\documentclass[onecolumn,showpacs,manuscript]{revtex4}
\usepackage{dcolumn}
\usepackage{amsmath}
\usepackage{amssymb}
\usepackage[dvips]{epsfig}  % to include PostScript figures
\makeatletter
\def\btt#1{\texttt{\@backslashchar#1}}%
\DeclareRobustCommand\bblash{\btt{\@backslashchar}}%
\makeatother

\begin{document}
%\begin{CJK*}{GBK}{song}
%\draft
\title{\bf Elastic properties of transition metal dioxides: \\XO$_{2}$(X=Ru, Rh, Os, Ir)}
\author{Yanling Li$^{1,2}$, Zhi Zeng$^{2}$}
\affiliation { 1.Department of physics, Xuzhou Normal University,
Xuzhou 221116, People's Republic of China\\ 2.Key Laboratory of
Materials Physics, Institute of Solid State Physics, Chinese
Academy of Sciences, Hefei 230031,People's Republic of China and
\\ Graduate School of the Chinese Academy of Sciences, Beijing
100049, People's Republic of China}

\date{\today}% It is always \today, today, but you may specify any date with \date.

\begin{abstract}
The elastic properties of rutile transition metal dioxides
XO$_{2}$ (X=Ru, Rh, Os, and Ir) are investigated using
first-principles calculations based on density functional theory.
Elastic constants, bulk modulus, shear modulus, and Young's
modulus as well as Possion ratio are given. OsO$_2$ and IrO$_2$
show strong incompressibility. The hardness estimated for these
dioxides shows that they are not superhard solids. The obtained
Debye temperatures are comparative to those of transition metal
dinitrides or diborides.

\end{abstract}

\pacs{61.50.-f, 62.20.-x, 71.15.Mb, 71.20.-b, 77.74.Bw, 78.20.Ga}% PACS, the Physics and Astronomy

\maketitle
\section{Introduction}
The rutile-type dioxides of transition metals (TMO$_2$) possess a
variety of electrical, mechanical, magnetic, and optical
properties. It's electronic structure and electrical properties are
widely studied due to their importance in technology. For example,
OsO$_2$ and RuO$_2$ can be used as a possible electrode material
in high-energy-density storage capacitors. Many of the TMO$_2$s,
such as RuO$_2$, IrO$_2$ and TiO$_2$, present a rich phase
diagram. They can transform to other structures under
high-pressure conditions. The first theoretical investigation on
the electronic structures of the rutile transition metal dioxides
(OsO$_2$, IrO$_2$, and RuO$_2$), were reported by Mattheiss using
a non-self-consistent
angmented-plane-wave-linear-combinations-of-atomic orbitals
method.\cite{Mattheiss} Subsequently, plenty of theoretical
researches, including on the electronic, optical and structural
properties were performed for rutile TMO$_2$. Xu \emph{et al}
studied the electronic and optical properties of NbO$_2$, RuO$_2$,
and IrO$_2$ using self-consistent calculations based on the linear
muffin-tin-orbital method and atomic sphere
approximation.\cite{Xu} Krasovska \emph{el al} calculated the
optical and photoelectron properties of RuO$_2$ using the \emph{ab
initio} self-consistent energy-band structure.\cite{Krasovska}
Recently, the electronic and optical properties of RuO$_2$ and
IrO$_2$ were reported by first-principles self-consistent
electronic structure calculations based on the full-potential
linearized plane wave method.\cite{Almeida}

In addition, the structural and mechanical properties of rutile
transition metal dioxides have been widely investigated
theoretically and experimentally. Structural phase transitions in
IrO$_2$\cite{Ono} and RuO$_2$\cite{Hugosson} were reported based
on first-principles calculations. Elastic properties of several
potential superhard RuO$_2$ phases were studied by
first-principles plane-wave pseudopotential and full-potential
linearized augmented plane-wave methods.\cite{Tse} Yen \emph{et
al} measured the first-order raman specta of OsO$_2$ at room
temperature.\cite{Yen} Recently, the lattice dynamics of RuO$_2$
was given by theory and experiment.\cite{Bohnen} However, to the
best of our knowledge, few research on the elastic properties of
IrO$_2$, OsO$_2$ and RhO$_2$ has been reported so far.

In this paper, we focus on investigating the elastic properties of
RuO$_2$, RhO$_2$, OsO$_2$ and IrO$_2$ using first-principles
calculations based on density functional theory (DFT). Firstly,
their structural parameters are optimized by first-principles
total energy calculations. Secondly, the elastic constants are
given by finite strain technique under the framework of linear
response theory. Then the incompressibility of these TMO$_2$s are
discussed. Finally, the hardness and Debye temperature are
estimated.

\section{Computational details}
All calculations are performed by the CASTEP code using \emph{ab
initio} pseudopotentials based on DFT. For the
exchange-correlation functional the generalized gradient
approximation (GGA), as given by Perdew, Becke, and Ernzerhof
(PBE), is applied. All the possible structures are optimized by
the BFGS algorithm which provides a fast way of finding the lowest
energy structure and supports cell optimization. In the
calculation, the interaction between the ions and the electrons is
described by using Vanderbilt's supersoft pseudopotential with the
cutoff energy of 380 eV. The Monkhorst-Pack
\emph{\textbf{k}}-point grid with a fine quality Brillouin zone
sampling of 2$\pi\times$0.04 \AA$^{-1}$ is used in the
calculations. In the geometrical optimization, all forces on atoms
are converged to less than 0.002 eV/\AA, all the stress components
are less than 0.02 GPa, and the tolerance in self-consistent field
(SCF) calculation is 5.0$\times$10$^{-7}$ eV/atom. The relaxation
of the internal degrees of freedom is allowed at each unit cell
compression or expansion. The elastic constants of TMO$_2$ are
obtained by using the finite strain technique. From the full
elastic constant tensor we determine the bulk modulus $B$, the
shear modulus $G$, the Young's modulus, $E$, and Possion's ratio
$\nu$ according to the Voigt-Reuss-Hill (VRH)
approximation.\cite{Hill}
%\begin{equation}\label{E-nu}\nonumber
%$E=\frac{9BG}{3B+G}$ , $\nu= \frac{3B-2G}{2(3B+G)}$.
%\end{equation}

\section{Results and discussions}

%\subsection{Structural property}
The tetragonal rutile structure belongs to space group $P42/mnm$.
There are six atoms per unit cell. Two metal atoms locate at (0,
0, 0) sites and four oxygen atoms position at ($u$, $u$, 0) sites,
where $u$ is the internal parameter. The metal atoms are
surrounded by six oxygen atoms at the corners of a slightly
distorted octahedron, while the three metal atoms coordinating
each of the oxygen atoms lie in a plane at the corners of a nearly
equilateral triangle. Equilibrium volumes, crystal constants, and
internal parameters are shown in Table I. It can be seen that all
the structural parameters optimized here agree well with previous
results. We find that there are very slight differences in volume
and internal parameter ($u$) between TMO$_2$s studied.

%\subsection{elastic property}
The calculated elastic constants $c_{ij}$ are given in Table II.
It is easy to see that these constants $c_{ij}$ satisfy the
Born-Huang stability criteria,
\begin{gather}
c_{ii}>0 (i=1,3,4,6), c_{11}>c_{12},\nonumber\\
c_{11}+c_{33}>2c_{13},2c_{11}+c_{33}+2c_{12}+4c_{13}>0,
\end{gather}
suggesting that they are mechanically stable. From the elastic
constants calculated above, we obtain the bulk modulus $B$, shear
modulus $G$, Young's modulus $E$, and Possion's ratio $\nu$ and
list them in Table II, which are important to understand the
elastic properties of TMO$_2$s. For RuO$_2$, our results agree
well with the available theoretical and experimental values,
indicating that PBE-GGA can be used to calculate the elastic
properties of TMO$_2$. OsO$_2$ and IrO$_2$ have the close bulk
moduli and Young's moduli, which are higher than those of RuO$_2$
and RhO$_2$. Higher bulk modulus indicates strong
incompressibility. From Table II, we conclude that IrO$_2$ and
OsO$_2$ are more difficult to be compressed than the two others,
meaning that IrO$_2$ and OsO$_2$ are hard solids. The
incompressibility of these TMO$_2$ can also be seen from relative
volume as a function of pressure (Fig. 1). While there isn't a
large difference in the shear modulus at four TMO$_2$s, it's
roughly the same. Also we note that there are few difference (13
GPa at most) between $c_44$ and shear modulus $G$ for four
dioxides.

Employing the correlation between the shear modulus and Vickers
hardness reported by Teter \cite{Teter}, the hardness of four
TMO$_2$s is estimated. By using the calculated $G$ values reported
in Table II and the correlation given by Teter we find a
theoretical Vickers hardness of approximately 14.1 GPa for RuO$_2$
, 12.3 GPa for RhO$_2$, 14.8 GPa for OsO$_2$ and 14.6 GPa for
IrO$_2$. It is well known that the hardness of superhard materials
should be higher than 40 GPa, so that TMO$_{2}$ are not superhard
materials.

Possion's ratio reflects the stability of a crystal against shear.
This ratio can formally take values between -1 and 0.5, which
corresponds respectively to the lower limit where the material
does not change its shape, and to the upper limit when the volume
remains unchanged. All the calculated Poisson's ratios (Table II)
are bigger than 0.25, which means that there are strong elastic
anisotropy in these TMO$_{2}$. In order to predict the brittle and
ductile behavior of solids, Puch introduced the ratio of the bulk
modulus to shear modulus of polycrystalline phases. A high (low)
\emph{B}/\emph{G} value is associated with ductility
(brittleness). The critical value which separates ductile and
brittle materials is about 1.75. The bigger difference obtained
between bulk and shear modulus indicates higher ductility of these
TMO$_{2}$ (Table II).

Now, we discuss the elastic anisotropy of TMO$_{2}$. For the
rutile structure, the directional bulk modulus along
crystallographic axis and the percent elastic anisotropy defined
by Chung and Buessem are discussed (Table III). From Table III, it
can be seen that the directional bulk modulus is larger along the
$c$ axis than that along the $a$$(b)$ axis, indicating that the
incompressibility along the $c$ axis is stronger than that along
the $a$ $(b)$ axis. This anisotropy of compression along different
lattice axis can be clearly seen from the increasing fractional
axis compression c/a versus pressure (Fig. 2). On the other hand,
the percentage of elastic anisotropy involves both the percentage
anisotropy in compressibility $A_{B}$ and in shear $A_{G}$. A
value of zero represents elastic isotropy and a value of 1 (100\%)
is the largest possible anisotropy. The calculated $A_{B}$ and
$A_{G}$ are listed in Table III. Obviously, these TMO$_{2}$s
possess strong shear anisotropy and weak bulk anisotropy.

Further, the Debye temperature $\Theta_{D}$ is also discussed
because the Debye temperature relates to many physical properties
of materials, such as specific heat, dynamical properties, and
melting temperature. It can be calculated from the average wave
velocity $\upsilon_{m}$ by the equation\cite{Anderson}
\begin{equation}
\Theta_{D}=\frac{h}{k}\left[\frac{3n}{4\pi}\left(\frac{\rho
N_{A}}{M}\right)\right]^{1/3}\upsilon_{m}
\end{equation}
where \emph{h} is Plank's constant, \emph{k} is Boltzmann's
constant, \emph{N$_{A}$} is Avogadro's number, $\rho$ is density,
\emph{M} is molecular weight and \emph{n} is the number of atom in
the molecule. $\upsilon_{m}$ is approximately given by
\begin{equation}
\upsilon_{m}=\left[\frac{1}{3}\left(\frac{2}{\upsilon^{3}_{t}}+\frac{1}{\upsilon^{3}_{l}}\right)\right]^{-1/3}
\end{equation}
where $\upsilon_{l}$ and $\upsilon_{t}$ are the transverse and
longitudinal elastic wave velocity of
 the polycrystalline material and are given by Navier's equation\cite{Schreiber}
\begin{equation}\label{tran}
\upsilon_{t}=\left(\frac{G}{\rho}\right)^{1/2}
\end{equation}
and
\begin{equation}\label{long}
\upsilon_{l}=\left(\frac{B+\frac{4G}{3}}{\rho}\right)^{1/2}.
\end{equation}
The calculated density, longitudinal, transverse, and average
elastic velocities and Debye temperatures $\Theta_{D}$ for four
TMO$_2$s are given in Table IV. Debye temperatures $\Theta_{D}$
of RuO$_2$ is comparative to the value of 866.4 $K$ in
ReB$_{2}$\cite{Hao} and ones of OsO$_2$ and IrO$_2$ is comparative
to the value of 691 $K$ in OsN$_{2}$.\cite{Wu}

%\subsection{Electronic property}

%-----------------------------------------------------Fig.------------------------

\section{Conclusion}
In summary, the elastic properties of transition metal dioxides
with rutile structure are investigated using first-principles
calculations under the framework of density functional theory
within generalized gradient density approximation. Elastic
constants, bulk modulus, shear modulus, and Possion's ratio are
given. OsO$_2$ and IrO$_2$ have stronger incompressibility than
RuO$_2$ and RhO$_2$. The estimated hardness is 14.1 GPa for
RuO$_2$, 12.3 GPa for RhO$_2$, 14.8 GPa for OsO$_2$ and 14.6 GPa
for IrO$_2$, which show that these dioxides are not superhard
solids. Debye temperatures calculated by elastic constants and
density are 846 K, 795 K, 674 K, 674 K for RuO$_2$, RhO$_2$,
OsO$_2$, and IrO$_2$, respectively.

\vspace{5 mm}
\section{Acknowledgement}
This work was supported by the National Science Foundation of
China under Grant Nos 10504036 and 90503005, the special Funds for
Major State Basic Research Project of China(973) under grant no.
2005CB623603, Knowledge Innovation Program of Chinese Academy of
Sciences, and  Director Grants of CASHIPS. Part of the
calculations were performed in the Shanghai Supercomputer Center.

\newpage

\noindent {\bf {\large {FIGURE CAPTIONS}}}

\vglue 1.0cm

\noindent {\bf {Fig.1:}} $V/V_0$ as a function of pressure for
seven TMO$_2$.

\vglue 1.0cm \noindent {\bf {Fig.2:}} c/a ratio as a function of
pressure for seven TMO$_2$.

\vglue 1.0cm

%\noindent {\bf {Fig.3:}} Internal free parameter $u$ in unit cell
%as a function of pressure.

\newpage
\noindent {\bf {\large {TABLE CAPTIONS}}}

\vglue 1.0cm \noindent {\bf {TABLE I:}} Equilibrium lattice
parameters,\emph{V}$_{0}$(\AA$^{3}$), \emph{a} (\AA), \emph{c}
(\AA), density $\rho$ (g/cm$^{3}$), and internal parameter $u$
along with the available experimental values.

\vglue 1.0cm \noindent {\bf {TABLE II:}} Zero-pressure elastic
constants \emph{c}$_{ij}$ (GPa), the isotropic bulk modulus
\emph{B} (GPa), shear modulus \emph{G} (GPa), Young's modulus
\emph{E} (GPa) and Possion's ratio $\nu$.

\vglue 1.0cm \noindent {\bf {TABLE III:}} The bulk modulus along
the crystallographic axes \emph{a}, and \emph{c} (\emph{B}$_{a}$
and \emph{B}$_{c}$) for TMO$_{2}$. Percent elastic anisotropy for
shear and bulk moduli \emph{A}$_{G}$ (in \%), \emph{A}$_{B}$ (in
\%) and compressibility anisotropy factors \emph{A}$_{B_{a}}$ and
\emph{A}$_{B_{c}}$. Here, $\emph{A}_{B_{a}}$
=$\frac{B_{a}}{B_{b}}$=1, $\emph{A}_{B_{c}}$
=$\frac{B_{c}}{B_{b}}$.

\vglue 1.0cm \noindent {\bf {TABLE IV:}} The longitudinal,
transverse, average elastic wave velocity ($\upsilon_l$,
$\upsilon_t$, and $\upsilon_m$ in $m/s$), and Debye temperature
$\Theta_{D}$ ($K$) at the theoretical equilibrium volume.

\newpage
%--------------------------------------Table I---------------------

%-----------------------------------------Table I---------------------------
\begin{table}[htbp]
\begin{center}
\caption{Li \textit {et al.}} \vspace{0.1cm}
\begin{minipage}{0.5\textwidth}
\begin{tabular}{ccccccc}
\hline \hline

 &\emph{V}$_{0}$  &\emph{a}  &\emph{c} &\emph{u}  &Reference\\
\hline

% TiO$_2$    &63.5799 &4.6348  &2.9598 &0.3051 &4.1725 &This \\
%                    &        &4.81   &5.64 &4.60           & &       && &[10]\\
% NbO$_2$    &71.4061 &4.9431  &2.9224 &0.2895 &5.8103 &This \\
%            &        &4.481   &2.992  &       &       &\cite{Xu}\\
%                     &        &4.7983  &&2.8077   &        &&&&[11]\\
% TaO$_2$    &73.1344 &5.0432  &2.8754 &0.2868 &9.6718 &This \\
 RuO$_2$    &63.1488 &4.5076  &3.1080 &0.3055  &This \\
            &        &4.4919  &3.1066 &0.306   &\cite{Mattheiss}\\
 RhO$_2$    &66.7652 &4.6051  &3.1482 &0.3067  &This             \\
 OsO$_2$    &64.3639 &4.4943  &3.1865 &0.3078  &This             \\
            &        &4.5003  &3.1839 &0.308   &\cite{Mattheiss} \\
 IrO$_2$    &67.4532 &4.5899  &3.2018 &0.3082  &This \\
            &        &4.4983  &3.1544 &0.307   &\cite{Mattheiss} \\
            &65.79   &4.541   &3.191  &0.3085  &\cite{Ono} \\
\hline \hline
\end{tabular}
\end{minipage}
\end{center}
\end{table}

\clearpage
\newpage
%------------------------------------------------Table II-------------------------
\begin{table}[htbp]
\begin{center}
\caption{Li \textit {et al.}}
\vspace{0.1cm}
\begin{tabular}{lcccccccccccc}
\hline \hline
       &       &\emph{c}$_{11}$  &\emph{c}$_{33}$  &\emph{c}$_{44}$  &\emph{c}$_{66}$ &\emph{c}$_{12}$
       &\emph{c}$_{13}$ &$B$    &$G$    &$E$   &$\nu$  &Ref.  \\
\hline

%&TiO$_2$  &265  &476  &117  &212  &162 &148  &208 &113 &288  &0.2694 &This  \\

%&NbO$_2$  &323  &360  &95   &243  &235 &161  &235 &102 &267  &0.3110 &This  \\

%&TaO$_2$  &379  &409  &98   &293  &283 &186  &275 &114 &299  &0.3185 &This  \\

&RuO$_2$  &303  &536  &117  &221  &240 &190  &260 &104 &275  &0.3239 &This          \\
&         &299  &558  &114  &227  &246 &199  &258 &    &     &       &\cite{Bohnen} \\
&         &     &     &     &     &    &     &270 &    &     &       &\cite{Hazen}  \\
&RhO$_2$  &268  &498  &103  &190  &216 &166  &232 &91  &242  &0.3263 &This  \\

&OsO$_2$  &337  &631  &109  &243  &270 &210  &292 &109 &291  &0.3338 &This  \\

&IrO$_2$  &344  &605  &117  &219  &274 &223  &299 &108 &290  &0.3381 &This  \\

\hline \hline
\end{tabular}
\end{center}
\end{table}

\clearpage
\newpage
%---------------------------------------------Table III-----------------------
\begin{table}[htbp]
\begin{center}
\caption{Li \textit {et al.}} \vspace{0.1cm}
\begin{tabular}{lcccccc}
\hline \hline
       &   &\emph{B}$_{a}$  &\emph{B}$_{c}$ &\emph{A}$_{B_{c}}$ &\emph{A}$_{B}$ &\emph{A}$_{G}$  \\
\hline

&RuO$_2$    &633   &1339  &2.1155 &1.66 &21.01 \\

&RhO$_2$    &560   &1219  &2.1751 &1.89 &22.24 \\

&OsO$_2$    &700   &1573  &2.2458 &2.02 &21.93 \\

&IrO$_2$    &718   &1600  &2.2284 &1.76 &18.83 \\

\hline \hline
\end{tabular}
\end{center}
\end{table}

%-----------------------

%----------------------------------------------Table IV-------------------------
\begin{table}[htbp]
\begin{center}
\caption{Li \textit {et al.}}
\vspace{0.1cm}
\begin{tabular}{lcccccc}
\hline \hline %%&\emph{A}$_{B_{a}}$ &\emph{A}$_{B_{c}}$
       &    &$\rho$        &$\upsilon_l$ &$\upsilon_t$ &$\upsilon_m$  &$\Theta_D$\\
\hline

%&TiO$_2$   &9279  &5213  &5801    &1134 \\

%&NbO$_2$   &7992  &4186  &4682    &881 \\

&RuO$_2$   &6.9995  &7548  &3852  &4316    &846 \\
&RhO$_2$   &6.7117  &7261  &3687  &4133    &795 \\
%&TaO$_2$  &9.6718  &6636   &3425  &3835    &715 \\
&OsO$_2$   &11.4687 &6180  &3086  &3463    &674 \\
&IrO$_2$   &11.0413 &6334  &3133  &3516    &674 \\

 \hline \hline
\end{tabular}
\end{center}
\end{table}
%-----------------------------------------------------------------

%--------------------------------------------------------
\clearpage
\newpage
%---------------------------------------Fig.1----------------------
\begin{figure}[htbp]
 \includegraphics[width=7.0 cm, angle=0,clip]{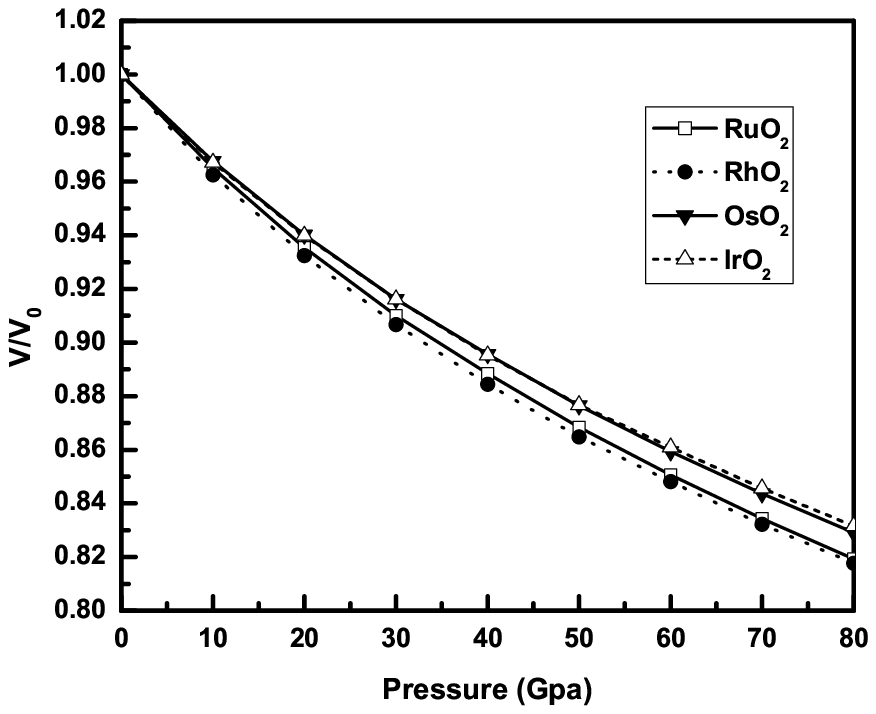}
 \caption{Li \textit {et al.}}
\end{figure}

\clearpage
\newpage
%---------------------------------------Fig.2----------------------
\begin{figure}[htbp]
\includegraphics[width=7.0 cm, angle=0,clip]{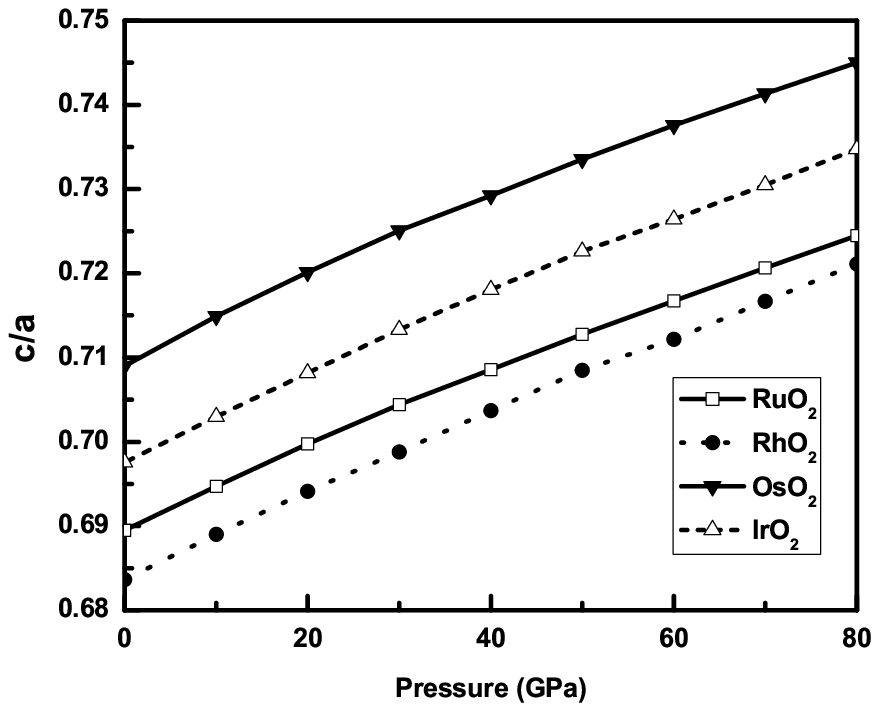}
\caption{Li \textit {et al.}}
\end{figure}
\clearpage

%-----------------------------------------Fig.3---------------------
%\begin{figure}[htbp]
%\includegraphics[width=7.0 cm, angle=0,clip]{u.eps}
%\caption{Li \textit {et al.}}
%\end{figure}
%\clearpage

%\end{CJK*}

\begin{thebibliography}{000}

\bibitem{Mattheiss}
L. F. Mattheiss, Phys. Rev. B {\bf13}, 2433 (1976).

\bibitem{Xu}
J. H. Xu, T. Jarlborg, and A. J. Freeman, Phys. Rev. B {\bf40},
7939 (1989).

\bibitem{Krasovska}
O. V. Krasovska, E. E. Krasovskii, and V. N. Antonov, Phys. Rev. B
{\bf52}, 11825 (1995).

\bibitem{Almeida}
J. S. de Almeida and R. Ahuja, Phys. Rev. B {\bf73}, 165102
(2006).

\bibitem{Ono}
S. Ono, J. P. Brodholt, and G. D. Price, J. Phys.: Condens. Matter
{\bf20}, 045202 (2008).

\bibitem{Hugosson}
H. W. Hugosson, G. E. Grechnev, R. Ahuja, U. Helmersson, L. Sa,
and O. Eriksson, Phys. Rev. B {\bf66}, 174111 (2002).

\bibitem{Tse}
J. S. Tse, D. D. Klug, K. Uehara, and Z. Q. Li, Phys. Rev. B
{\bf61}, 10029 (2000).

\bibitem{Yen}
P. C. Yen, R. S. Chen, Y. S. Huang, C. T. Chia, R. H. Chen, and K.
K. Tiong, J. Phys.: Condens. Matter {\bf15}, 1487 (2003).

\bibitem{Bohnen}
K. -P. Bohnen, ei al, Phys. Rev. B {\bf75}, 092301 (2007).

\bibitem{Hill}
R. Hill, Proc.Phys.Soc.London {\bf 65}, 349 (1952).

\bibitem{Hazen}
R. M. Hazen and L. W. Finger, J. Phys. Chem. Solids {\bf42}, 143
(1981).

\bibitem{Teter}
D. M. Teter, MRS. Bull. {\bf 23}, 22 (1998).

\bibitem{Anderson}
O. L. Anderson, J. Phys.Chem. Solids {\bf 24}, 909 (1963).

\bibitem{Schreiber}
E. Schreiber, O. L. Anderson, and N. Soga, Elastic Constants and
their Measurements (McGraw-Hill, New York, 1973).

\bibitem{Hao}
X. Hao, Y. Xu, Z. Wu, D. Zhou, X. Liu, X. Cao, and J. Meng,
Phys.Rev.B {\bf 74}, 224112 (2006).

\bibitem{Wu}
 Z. Wu, X. Hao, X. Liu, and J. Meng, Phys.Rev.B {\bf 75}, 054115 (2007).

\end{thebibliography}
\end{document}